\documentclass[preprint,nofootinbib,preprintnumbers,floats,aps]{revtex4}

\usepackage{amsmath}
\usepackage{amssymb}
\usepackage{graphicx}
\usepackage{epsfig}
\usepackage{bm}
\usepackage{subfigure}

\newcommand{\be}{\begin{equation}}
\newcommand{\ee}{\end{equation}}
\newcommand{\bea}{\begin{eqnarray}}
\newcommand{\eea}{\end{eqnarray}}
\newcommand{\ben}{\begin{enumerate}}
\newcommand{\een}{\end{enumerate}}
\newcommand{\bit}{\begin{itemize}}
\newcommand{\eit}{\end{itemize}}

\newcommand{\vv}{\boldsymbol}					

\newcommand{\bert}{\raise-0.45mm\hbox{\Large$\Box$}}

\begin{document}

\preprint{MIT-CTP-4025}
\preprint{0904.0453 [gr-qc]}

\title{Constraints on emergent gravity}
 
\author{Alejandro Jenkins}\email{ajv@mit.edu}

\affiliation{Center for Theoretical Physics,
Laboratory for Nuclear Science and Department of Physics,
Massachusetts Institute of Technology, Cambridge, MA 02139, USA
\vskip 0.5cm
}

\begin{abstract}

\vskip 0.5cm

In this essay we review the central difficulty in formulating a viable quantum field theory in which gravity is emergent at low energies, rather than mediated by a fundamental gauge field.  The Weinberg-Witten theorem forbids spin-2, massless modes from carrying Lorentz covariant stress-energy.  In General Relativity the stress-energy is not covariant because it violates a gauge invariance, but a gravitational theory without fundamental spin-2 gauge invariance must either lack a stress-energy operator or have a non-relativistic graviton.  The latter option is incompatible with the principle of equivalence, though such theories are not necessarily ruled out at low energies.

\vskip 1.5cm

\begin{center}
{\it Essay written for the Gravity Research Foundation \\
2009 Awards for Essays on Gravitation} \\
Awarded honorable mention
\end{center}

\end{abstract}

\maketitle

\section{Introduction}
\label{intro}

The classical theory of General Relativity (GR) emerges at low energies from a relativistic quantum field theory (QFT) with an interacting, massless, spin-2 particle (the graviton) \cite{lineartoGR,DeserlineartoGR,weinbergGR,boulwaredeser}.  The graviton's coupling has dimensions of mass$^{-1}$ (in units in which $c=\hbar =1$), which implies that perturbative scattering amplitudes grow with energy.  The quantized theory therefore appears to be non-renormalizable (for a review see, {\it e.g.}, \cite{divergences}).

String theory solves the problem of non-renormalizability and is widely seen as the most promising proposal for a consistent quantum theory of gravity, but its implications are far from being fully understood.  A conceivable alternative is to take the graviton to be an emergent, low energy degree of freedom, like the pion in the case of the strong interactions.  In this essay we review the reasons why this idea is difficult to implement, though a viable theory of gravity along these lines is not impossible.

\section{Lorentz and gauge symmetries}
\label{lorentz}

Local gauge invariance is distinct from the other physical symmetries in that, rather than relating different physical setups, it declares that two different mathematical descriptions correspond to {\it the same physical setup}.  This redundancy is a necessary feature of the relativistic description of massless particles with spin $j > 1/2$.  (For a full treatment of this interpretation of gauge invariance in the spin 1 case, see \cite{weinbergFT}.)

For spin 2, a Lorentz covariant tensor needs five polarizations, but a massless particle has only two.  Under a Lorentz transformation $\Lambda$, a massless spin-2 field $h_{\mu \nu}$ therefore transforms non-covariantly, as
\be
h_{\mu\nu} (x) \to \Lambda_\mu^{~\rho} \Lambda_\nu^{~\sigma} h_{\rho \sigma} (\Lambda x)
+ \partial_\mu \xi_\nu(x, \Lambda) + \partial_\nu \xi_\mu(x, \Lambda)~.
\label{gauge}
\ee
Maintaining physical Lorentz symmetry requires that we mod out graviton states by all gauge terms of the form $\partial_\mu \xi_\nu + \partial_\nu \xi_\mu$, and the graviton's interactions must be such that these terms drop out of the expressions for observable quantities.  In classical GR, this gauge invariance becomes general coordinate invariance \cite{weinbergGR}.

Weinberg showed that Lorentz invariance in the QFT requires that the graviton couple universally to the conserved energy-momentum \cite{infrared}.  Otherwise, a Lorentz transformation will introduce a pure gauge term in the amplitude for soft graviton emission, leading to an IR divergence (see also  Sec. 13.1 in \cite{weinbergFT}).  Thus, the graviton must couple to the Noether stress-energy
\be
T^\mu_{~\nu} =
\frac{\partial {\cal L}} {\partial (\partial_\mu g_{\alpha \beta} )} (\partial_\nu g_{\alpha \beta})
- \delta^\mu_{\phantom a \nu} {\cal L}~,
\label{gravityT}
\ee
where $g_{\mu \nu} = \eta_{\mu \nu} + h_{\mu\nu}$ to linear order.  But the Einstein-Hilbert action contains terms linear in second derivatives of the metric $g_{\mu \nu}$.  To define the stress-energy by the Noether procedure, we must integrate these by parts and drop the boundary term, destroying the gauge invariance (see, {\it e.g.}, Ch. 31 in \cite{dirac}). This reflects the fact that the conserved energy-momentum cannot be defined in a coordinate invariant way in GR and is therefore not a local observable.\footnote{The fact that the conserved energy-momentum is not a local observable in GR follows from the principle of equivalence: there is always a frame of reference in which the gravitational field vanishes locally.}  Meanwhile, the gauge invariant 
\be
\Theta^{\mu  \nu} = \frac{1}{\sqrt{-g}} \frac{\delta S_{\rm matter}}{\delta g_{\mu \nu}}
\label{matterT}
\ee
measures the stress-energy of everything {\it except} gravity, and is therefore not conserved (see, {\it e.g.}, Sec. 7.6 in \cite{gravity}).  For spin 1, all of this has a close parallel in Yang-Mills theories, where the conserved current is not gauge invariant because the gauge field is charged.

\section{Weinberg-Witten theorem}
\label{WW}

Let $| p \rangle$ and $| p' \rangle$ be one-particle, spin-2, massless states ---labeled by their 4-momenta---  with the same Lorentz-invariant helicity $\pm 2$.  The Weinberg-Witten theorem \cite{weinbergwitten} establishes that if the matrix elements $\langle p' | T^{\mu \nu} | p \rangle$ are Lorentz covariant, then
\be
\lim_{p' \to p} \langle p' | T^{\mu \nu} | p \rangle = 0~,
\label{matrix}
\ee
which prevents the particle from carrying observable energy-momentum (for detailed reviews of the Weinberg-Witten theorem, see \cite{thesis,loebbert}).  Therefore, any theory with a non-trivial, massless tensor mode in its spectrum must
\ben
\item[(a)] lack a stress-energy operator, or
\item[(b)] have a fundamental gauge invariance, under which the stress-energy is {\it not} invariant, making the matrix elements $\langle p' | T^{\mu \nu} | p \rangle$ mathematically non-covariant by Eq. (\ref{gauge}), or 
\item[(c)] have an emergent space-time, distinct from the background space-time, in which the emergent graviton propagates and has an emergent gauge invariance,\footnote{It does not make sense to consider an emergent, spin-2 gauge invariance in the {\it background} space-time, because such an invariance prevents gravitational energy from being locally observable, and is therefore at odds with the separation of energy scales needed to formulate an effective field theory of emergent gravity.  On this point, see also \cite{polchinski,sidneyfest}.} 
\item[(d)] have non-relativistic gravitons.
\een

First-quantized string theory falls under category (a), because there is no consistent, off-shell definition of the string action $S$ in the background space-time with metric $g_{ab}$, so that the object
\be
T^{a b} = \frac{1}{\sqrt{-g}} \frac{\delta S}{\delta g_{a b}}
\label{stringT}
\ee
is undefined.  Other intrinsically nonlocal theories, such as Sundrum's ``fat gravitons'' \cite{fatgravity} also fall into this category (on this point, see also \cite{take}).  GR (and, presumably, string field theory) fall under category (b).  The AdS/CFT correspondence falls under category (c).  Proposals for emergent gravitons motivated by condensed matter physics usually fall under category (d): for some recent examples, see \cite{zhanghu,kraustomboulis,wetterich,bob,horava}.

\section{Effective graviton coupling}
\label{coupling}

A complete theory of emergent gravity should account for the non-gravitational physics ({\it i.e.}, the Standard Model) as well as for a gravity-like interaction at low energies.  Since the Standard Model is known to be Lorentz invariant to very high accuracy (see, {\it e.g.}, \cite{mattingly}), the underlying theory should be well described by a relativistic QFT, at least up to the TeV scale.  However, if that theory has a gauge invariant, conserved stress-energy,\footnote{Yang-Mills theories like the Standard Model have such a stress-energy: the Belinfante tensor (see, {\it e.g.}, Sec. 7.4 in \cite{weinbergFT}), which may also be expressed by Eq. (\ref{matterT}) evaluated at $g_{\mu \nu} = \eta_{\mu \nu}$.} then the Weinberg-Witten theorem requires that the emergent graviton's energy-momentum be {\it physically} Lorentz non-covariant.

\begin{figure} [t]
\begin{center}
\includegraphics[scale=1.5]{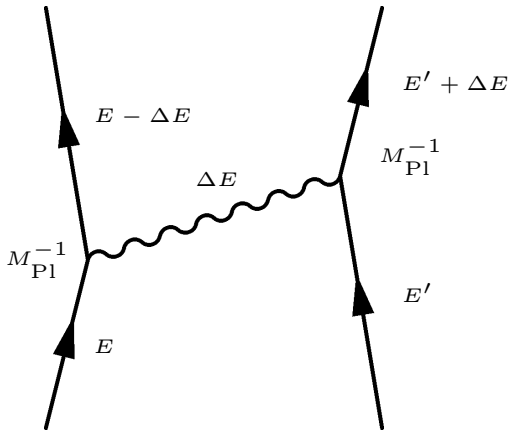}
\end{center}
\caption{\small Diagram for the exchange of energy $\Delta E$ via an emergent graviton.  If $E, E' \sim \mu$, then by dimensional analysis we typically expect $\Delta E \sim \mu^2 / M_{\rm Pl}$.  Meanwhile, by Eq. (\ref{nonrelgraviton}), $\Delta E \sim \bar L \mu$.}
\label{effectivegraviton}
\end{figure}

Let $\bar L$ be the order of the gravitational Lorentz violation, {\it i.e.}, the fraction of the gravity sector's energy-momentum ---defined in the appropriate preferred frame of reference--- which does not transform covariantly.  For instance, a graviton with a dispersion relation $E = v | \vv p |$ would have $\bar L \sim (c - v) / c$, where $c$ is the limiting speed in the Lorentz transformations.  In the preferred frame, Eq. (\ref{matrix}) becomes
\be
\lim_{p' \to p} \int d^3 x \, \langle p' | T^{00} | p \rangle = p^0 \sim \bar L \cdot \mu~,
\label{nonrelgraviton}
\ee
where $\mu$ is the energy scale of the interactions.  In other words, the Weinberg-Witten theorem implies that the fraction of the conserved energy {\it that can couple like a graviton} is of order $\bar L$.  Since the energy $p^0$ carried by a graviton is of order $\mu^2 / M_{\rm Pl}$, where $M_{{\rm Pl}}^{-1}$ is the effective gravitational coupling to matter (see Fig. \ref{effectivegraviton}), we conclude that
\be
\frac{\mu}{M_{\rm Pl}} \sim \bar L~.
\label{suppression}
\ee
For instance, if the emergent graviton is a Goldstone boson of spontaneous Lorentz violation, as in \cite{kraustomboulis,wetterich,bob}, $\bar L \to 0$ corresponds to the restoration of the symmetry and the disappearance of the gravitational degrees of freedom ($M_{\rm Pl} \to \infty$).  In Ho\v{r}ava's implementation of emergent gravity \cite{horava}, the graviton becomes relativistic only as $\mu \to 0$.

\section{Principle of equivalence}
\label{EP}

As we pointed out in Sec. \ref{lorentz}, the fact that in GR gravity couples universally to the conserved stress-energy follows from invariance of observables under gauge transformations
\be
h_{\mu \nu} \to h_{\mu \nu} + \partial_\mu \xi_\nu + \partial_\nu \xi_\mu~.
\label{gauge2} 
\ee
This universality ---the principle of equivalence--- is enforced in QFT by the Ward identity of the spin-2 gauge invariance \cite{gravityward}.  Lorentz violation breaks this gauge invariance and consequently also the principle of equivalence (see, {\it e.g.}, \cite{mattingly}).\footnote{We could, of course, replace Lorentz symmetry by a different symmetry, {\it e.g.} by working around de Sitter space-time \cite{deSitter}.  The Standard Model would respect de Sitter (rather than Lorentz) invariance and the de Sitter analog of the Weinberg-Witten theorem would require either a local gauge invariance or physical non-covariance of the graviton's energy-momentum.  The latter would be incompatible with equivalence.}

As Witten has emphasized in \cite{sidneyfest}, the key problem for emergent gravity is that exact equivalence implies a general coordinate invariance that prevents gravity from having a {\it physical} cutoff scale in that space-time, because a cutoff cannot be defined in a coordinate-invariant way.  Equivalence has been experimentally verified at the level of $\sim 10^{-13}$, but only in the coupling of gravity to matter \cite{poem}.  It is possible that some new dynamical principle might give the non-relativistic graviton a universal or quasi-universal couplings to non-gravitational stress-energy.  If Lorentz symmetry is broken {\it only} by gravity, a spurion analysis shows that the graviton's {\it self-coupling} will violate equivalence at order $\bar L$,  but there are no direct observational constraints on this (see, {\it e.g.}, Sec. 3.1 of \cite{gravity}, where it is referred to as violation of the ``very strong principle of equivalence'').\footnote{Vector-tensor theories of gravity \cite{vectortensor} are examples of theories that break Lorentz symmetry and equivalence only in the gravity sector.  Their gravitational vector field can be interpreted as a spurion of the breaking of Lorentz invariance \cite{graesser}.  For constraints on these theories, see \cite{jacobson}.}

Particles traveling faster than the non-relativistic graviton's speed $v$ would lose energy by gravitational \v{C}erenkov radiation.  Limits on this effect from high-energy cosmic rays indicate that $(c - v) / c \lesssim 10^{-15}$ \cite{nelson}, which may generically constrain emergent gravity much more than tests of equivalence.\footnote{These bounds are absent if the graviton is superluminal ($v > c$), which would require abandoning Lorentz symmetry at high energies \cite{jacobson}.}  Without exact equivalence, however, gravity at higher energies may look very different from GR, with consequences for the cosmology of the early universe, black holes,\footnote{Gravitational Lorentz violation may conflict with the laws of black hole thermodynamics \cite{LVbh}.  There are no direct constraints of such effects, but they could raise problems for the consistency of quantum gravity.} {\it etc.}

\section{Summary}
\label{summary}

In local field theory, if the graviton is not a gauge particle, then it must be non-relativistic.  In that case, the order of the gravitational Lorentz violation will also be the order of the violation of the principle of equivalence, as well as the order by which the graviton's effective coupling is suppressed with respect to the energy scale of the interactions.  Such theories are not necessarily ruled out, at least at low energies.  Whether the apparent cost in elegance of working out a viable theory of this sort outweighs the potential advantage of addressing the problem of quantizing gravity within the context of ordinary field theory, seems open to debate.

\begin{acknowledgements}
We thank Ted Jacobson and Bob McElrath for valuable feedback on the draft of this essay, and Nabil Iqbal, Vijay Kumar, Florian Loebbert, John McGreevy, Lubo\v{s} Motl, Takemichi Okui, T. Padmanabhan, Matthew Schwartz, Mark Wise, and Barton Zwiebach for discussions.  This work was supported in part by the U.S. Department of Energy under contract DE-FG03-92ER40701.
\end{acknowledgements}


\bibliographystyle{aipprocl}   


\end{document}